\documentclass[final,5p,times,twocolumn]{elsarticle}

\usepackage{hyperref}
\usepackage{graphicx}

\usepackage{amssymb,amsmath,amsfonts}
\usepackage{amsthm}

 \newcommand{\bfE}{\mathbf{E}}
\newcommand{\bE}{\mathbf{E}}
\newcommand{\bfB}{\mathbf{B}}
 \newcommand{\bfA}{\mathbf{A}}
\newcommand{\bB}{\mathbf{B}}
\newcommand{\bfJ}{\mathbf{J}}
\newcommand{\bJ}{\mathbf{J}}
\newcommand{\bfv}{\mathbf{v}}

\newcommand{\bfx}{\mathbf{x}}
\newcommand{\bv}{\mathbf{v}}
\newcommand{\bx}{\mathbf{x}}

\newcounter{bla}

\journal{Journal of Computational Physics}

\begin{document}

\begin{frontmatter}



\title{Exactly Energy Conserving Semi-Implicit  Particle in Cell Formulation}


\author[a]{Giovanni Lapenta\corref{author}}

\cortext[author] {Corresponding author.\\\textit{E-mail address:} giovanni.lapenta@kuleuven.be}
\address[a]{Department of Mathematics, KU Leuven, University of Leuven, Belgium}

\begin{abstract}
We report a new particle in cell (PIC) method based on the semi-implicit approach. The novelty of the new method is that unlike any of its semi-implicit predecessors at the same time retains the explicit computational cycle and conserves energy exactly. Recent research has presented fully implicit methods where energy conservation is obtained as part of a non linear iteration procedure. 
The new method (referred to as Energy Conserving Semi-Implicit  Method, ECSIM), instead, does not require any non-linear iteration and its computational cycle is similar to that of explicit PIC. 

The properties of the new method are: i) it conserves energy exactly to round-off for any time step or grid spacing; ii) it is unconditionally stable in time, freeing the user from the need to resolve the electron plasma  frequency and allowing the user to select any desired time step; iii) it eliminates the constraint of the finite grid instability, allowing  the user to select any desired resolution without being forced to resolve the Debye length; iv) the particle mover has a computational complexity identical to that of the explicit PIC, only the field solver has an increased computational cost. 

The new ECSIM is tested in a number of benchmarks where  accuracy and computational performance are tested.
\end{abstract}


\end{frontmatter}

\section{Introduction}
\label{introduction}
The particle in cell (PIC) method has affirmed itself as one of the most widely used numerical methods to model plasmas at the kinetic level \cite{birdsall-langdon,hockney-eastwood,GrigoryevPIC}. The method is often selected for its simplicity that makes any new user or developer readily understand the spirit and implementation. The simplicity is also a great advantage when considering computer science issues relative to the implementation of PIC. Widely different approaches can be quickly implemented to test new software practices and new computer architectures \cite{markidis2005implementation, decyk2011adaptable, verleye2013implementation, wuyts2015helsim}.

While the PIC method has proven  over the last several decades its ability of giving the right answers in many scientific and engineering problems, a number of further improvements are being studied. Prominent among them is the conservation of energy. 

The most common implementation of the PIC method is the classic explicit PIC presented in every textbook. The idea is to break the link between particles and fields for the duration of one time step. In each time step, the particles are advanced in the old fields and afterwards the fields are advanced with the particle information just updated. This approach is tremendously simple and successful but at one cost: it does not conserve energy. In fact energy tends to grow in time. This limitation is  overcome in practice by judiciously choosing the time step. With sufficiently small, sometimes really very small~\cite{lapenta2011particle}, time steps, energy can be conserved to an acceptable degree.


Recent research has focused on developing a new approach to PIC where energy is exactly conserved\cite{markidis2011energy,chen2011energy,taitano2013development} (we refer to this approach ECPIC in the following). This important result is obtained at the cost of making the PIC scheme fully implicit, meaning that the particle equations and the field equations have to be solved together, coupled via a non-linear Newton or Picard iteration. Energy conservation is in principle exact and its accuracy in practice is controlled by the tolerance of the iterative Newton-Krylov solver used.

The question we ask here is: can energy conservation be achieved in a scheme that retains the feel of explicit PIC avoiding the non-linear iteration? We ask then if one can retain the marching order of explicit PIC: advancing for one time step  the particles first and then the fields, then the particles again, without any iteration. 

The answer we find is yes. We can, to our knowledge for the first time, design a method that retains that very desirable simplicity of explicit schemes and yet conserves energy exactly. A way to reach this goal is to use the new implementation of the semi-implicit PIC method presented here, called energy conserving semi-implicit  method (ECSIM). 

 Semi-implicit methods  retain some of the particle coupling in the field equations through extra terms in the Maxwell's equations that provide a linear approximation to the particle response. The method still advances in its computational cycle as an explicit method but each step in the cycle is somewhat  more complex than in the explicit PIC method.  Two types of semi-implicit methods have been reported in the literature:  the direct implicit method (DIM), developed at Lawrence Livermore National Laboratory and implemented in codes such as AVANTI \cite{Hewett:1987,directimplicit} and  LSP \cite{welch2004implementation} and the implicit moment method (IMM) developed at Los Alamos National Laboratory\cite{brackbill-forslund,brackbill2014multiple} and used in a family of codes originating with Venus~\cite{brackbill-forslund}, followed by Celeste3D~\cite{vu,Lapenta:2006}, Parsek2D~\cite{parsek} and now used in Parsek2D-MLMD~ \cite{beck2014multi,innocenti2015introduction} and iPic3D~\cite{ipic3d}.

Compared with these previous semi-implicit methods, we modify significantly the particle mover and the mathematical procedure used to derive the field equations. 
The mover is replaced with a new mover that does not require any inner iteration and is in its practical implementation of the same level of complexity as the mover of explicit PIC. The field solver on the other hand is significantly more advanced than in the standard explicit methods requiring a significant amount of extra CPU effort. But it remains formally simple, similar to previous semi-implicit methds. Recalling that the cost of the PIC method is primarily coming from the particle mover, the new method has much potential in terms of computing performance.

The previous semi-implicit methods had many good properties of stability and ability to handle multiple time scale problems~\cite{brackbill2014multiple,brackbill-forslund,directimplicit}. But they did not conserve energy \cite{brackbill2015energy, directimplicit-optimisation}. The new method reported here, instead conserves energy exactly. 

The practical implication of full energy conservation will be fully assessed in real world applications that we hope future research using the new method will attempt. We report here results based on a simple but complete MATLAB implementation. Using it we show the following points: 
\begin{enumerate}
\item The method indeed conserves energy exactly as the mathematical proof we provide promises. We use a direct solver so the conservation is exact to round-off. When iterative linear solvers are used, the accuracy of energy conservation would be controlled by the tolerance of the linear solver. 
\item The so-called finite grid instability that in explicit PIC forces users to resolve the Debye length can be completely overcome. We show examples where the grid spacing is 16 orders of magnitude larger than the Debye length and the new method continues to remain stable and well behaved. 16 orders is not a limit, we just stopped out of fatigue. There is no reason to believe the system size could not be increased further while still retaining stability. 
\item The anisotropic heating and cooling of standard PIC or even of semi-implicit PIC is completely avoided, just as in fully implicit PIC
\item The complexity (in terms of number of cpu operations) of the new method is of the same order as in previous semi-implicit PIC
\end{enumerate}

Below, Section 2 is the central section of the present manuscript and derives the new method, the ECSIM. Section 3 reports the formal proof of energy conservation, exact and to round off. Section 4 provides several benchmarks that confirm the ability of the new method to conserve energy exactly.  Section 5 focuses on some additional properties relative to the ability of ECSIM to eradicate completely the grid spacing limitation of the finite grid instability and the unphysical  and anisotropic heating/cooling of species. Section 6 draws the conclusions and suggest further needed research.

The appendices report the details of the stability analysis of the new method (Appendix A),  the details of the spatial discretization used in the tests reported in Section 4  and 5 (Appendix B) and the order of accuracy in space and time of the specific implementation used here (Appendix C). Appendix D summarizes all the equations comprising the computational cycle of the new method, providing a concise summary to the interested code developer. 

\section{Energy Conserving Moment Implicit Method}
Our goal is to derive a variant of the semi-implicit method that conserves energy exactly, borrowing this important property from fully implicit ECPIC, but avoiding the need for a non-linear iteration between particles and fields. Like previous semi-implicit methods, we still rely on linear field-particle coupling to retain the explicit nature of the PIC computational cycle. 
 
Our starting point is a mover that combines the DIM $D_1$ scheme \cite{Hewett:1987} with the IMM and ECPIC $\theta$-scheme \cite{brackbill-forslund}. We advance the particle position as in the $D_1$ scheme\cite{directimplicit}, but advance the velocity according to the $\theta$ scheme \cite{vu}.

This is the crucial innovation that allows energy conservation and complete stability of the method. We evaluate the fields at the time staggered particle position $\bfx_p^{n+1/2}$ known explicitly from the time staggering of the leap-frog but we use the electric field at the $\theta$ time level: $\bfE^{n+\theta}$ exactly as in the standard IMM. The mover has the same stability properties of the IMM (Appendix A) but requires no predictor-corrector  iteration between the particle position and particle velocity. This feature is what allows us to write the scheme to be exactly energy conserving. The order of accuracy of the mover is second order in the coupling between the particle position and velocity and in the coupling with the electric field (if $\theta=1/2$) but it is first order in the coupling with the magnetic field (Appendix C reports more details).

The proposed new mover is:
\begin{equation}
\begin{array}{c}
\displaystyle \bfx_p^{n+1/2}=\bfx_p^{n-1/2} + \Delta t \bfv_p^{n}\\ \\
\displaystyle \bfv_p^{n+1}=\bfv_p^{n} + \frac{q_p \Delta t}{m_p} \left(  \bfE^{n+\theta}(\bfx_p^{n+1/2}) + \overline{\bfv}_p\times \bfB^{n}(\bfx_p^{n+1/2}) \right)
\end{array}
\label{thetaECSIM}
\end{equation}
where $\overline{\bfv}_p= (\bfv_p^{n+1}+ \bfv_p^{n})/2$.
As can be seen the first equation is the same as in the $D_1$ scheme (but in this regard, identical also to the simple leap-frog scheme), it can be solved using the known velocity from the previous time step. The second equation is identical to that of the $\theta$-scheme but the electric and magnetic fields are computed at the known position $\bfx_p^{n+1/2}$ rather than at the unknown position $\overline{\bfx}_p$ that in the standard IMM requires the predictor-corrector iteration.
These two positions are conceptually the same, they express the particle position at the mid-time between the old and new evaluations of the velocity. But one is computed explicitly while the other is computed as part of a predictor-correctors iteration. Both are second order accurate but the scheme in eq. (\ref{thetaECSIM}) is simpler to compute. 

For stability what counts is how the acceleration is computed. The magnetic field term is unconditionally stable even in the standard Boris mover~\cite{hockney-eastwood}. The term possibly giving a stability constraint is the electric field term. Similarly, the magnetic field does not do any work on the particles and causes no worry in energy conservation, only the electric field term does. 
For these reasons the force term is written using the magnetic field at the initial time level $\bfB^{n}(\bfx_p^{n+1/2})$ but the electric field is written at the advanced intermediate level $\bfE^{n+\theta}(\bfx_p^{n+1/2})$.
This same approach is used in the standard $\theta$ scheme so the stability analysis remains the same as in the $\theta$ scheme and the method is unconditionally stable, as we show in  \ref{stability-mover}.

The fields at the particle positions are computed by interpolation:
\begin{equation}
\bfE_p^{n+\theta}=\bfE^{n+\theta}(\bfx_p^{n+1/2})=\sum_g \bfE_g^{n+\theta} W(\bfx_p^{n+1/2}-\bfx_g)
\label{interpE}
\end{equation}
where we have assumed that the field equations are discretized on a grid with a generic index $g$ (that can be for example the ensemble of the indices $i$, $j$, $k$ in 3D Cartesian grids). Appendix B reports the actual discretization used in the examples below. A similar expression holds for the magnetic field:
 \begin{equation}
\bfB_p^{n+\theta}=\bfB^{n+\theta}(\bfx_p^{n+1/2})=\sum_g \bfB_g^{n+\theta} W(\bfx_p^{n+1/2}-\bfx_g)
\label{interpB}
\end{equation}
A short hand notation has been introduced: $\bfB_p^{n+\theta}=\bfB^{n+\theta}(\bfx_p^{n+1/2})$ and $\bfE_p^{n+\theta}=\bfE^{n+\theta}(\bfx_p^{n+1/2})$.

The interpolation function $W$ can be chosen in many different ways but we consider here the case of b-splines of order $\ell$ \cite{bspline} and we use the same interpolation functions for both electric and magnetic field (some methods use different interpolations for the two).  For example in 3D Cartesian grids this becomes:
\begin{equation}
W(\bfx_p-\bfx_g)=b_\ell(x_p-x_g)b_\ell(y_p-y_g)b_\ell(z_p-z_g)
\end{equation}
We will use b-spline of order one. This expression reduces trivially in 1D for the examples reported below.

For the Maxwell's equation we use the same discretization as in the standard IMM \cite{ipic3d}.  The two curl Maxwell equations are discretized in time with another $\theta$-scheme:
\begin{equation}
\begin{array}{ccc}
\displaystyle \nabla_g \times \bfE^{n+\theta} + \frac{1}{c} \frac{\bfB^{n+1}_g-\bfB^n_g}{\Delta t} =0\\ \\
\displaystyle \nabla_g \times \bfB^{n+\theta} - \frac{1}{c} \frac{\bfE^{n+1}_g-\bfE^n_g}{\Delta t} =\frac{4\pi}{c} \overline{\bfJ}_g
\end{array}
\label{maxwell-discrete}
\end{equation}
The spatial operators in eq. (\ref{maxwell-discrete}) are discretized on a grid. Consistent with the notation introduced earlier, we index the grid elements generically with the index $g$. $\nabla_g$ is a shorthand for the spatial discretization used. For example in the IMM family of codes the discretization used is different from the Yee scheme \cite{yee1997finite} and  locates on the cell centers  the magnetic field and on the cell vertices for the electric field~\cite{sulsky1991numerical,ipic3d,lapenta2012particle}. Appendix B reports the actual discretization used in the examples below. But all the derivations below are not critically dependent on which spatial discretization is used.

The set of Maxwell's and Newton's equations are coupled. Eq. (\ref{maxwell-discrete}) requires the new particle velocities to compute the current and eq.(\ref{thetaECSIM}) requires the new advanced electric field to move the particles. In the spirit of the semi-implicit method, we do not want to solve two coupled sets with a single non-linear iteration and find instead a way to extract analytically from the equations of motion the information needed for computing the current without first moving the particles. In previous semi-implicit methods this is done via a linearization procedure. The new mover used here allows us, instead, to derive the current rigorously without any approximation. 

Form its definition, the current   for each species in each grid location is:
\begin{equation}
\overline{\bfJ}_{sg}=\frac{1}{V_g}\sum_{p \in s} q_p \overline{\bfv}_p W(\bfx_p^{n+1/2}-\bfx_g)
\label{currentECSIM}
\end{equation}
where the summation is over the particles of the same species, labeled by $s$.

With easy vector manipulations~\cite{vu}, the velocity equation can be rewritten in the equivalent  form: 
\begin{equation}
\overline{\bfv}_p=\widehat{\bv}_p+
\beta_s\widehat{\bE}_p
\label{theta-rotated}
\end{equation}
where hatted quantities have been rotated by the magnetic field:
\begin{equation}
\begin{array}{c}
 \widehat{\bv}_p = {\alpha}^n_p  \bv^n_p \\ \\
\widehat{\bE}_p = {\alpha}^n_p 
\bE_p^{n+\theta} 
\end{array}
\label{hatted}
\end{equation}
via a rotation matrix ${\alpha}_p^n$ 
defined as:
\begin{equation}
{\alpha}_p^n =  \frac{1}{1+(\beta_s B_p^{n})^2}
\left(\mathbb{I}-\beta_s \mathbb{I} \times \bB_p^n +\beta_s^2
\bB_p^n \bB_p^n \right)
\label{alpha}
\end{equation}
where $\mathbb{I}$ is the dyadic tensor (matrix with diagonal of 1) and $\beta_s=q_p \Delta t/2m_p$ (independent of the particle weight
and unique to a given species).  The elements of the rotation matrix are indicated as 
${\alpha}^{ij,n}_p $ with label $i$ and $j$ referring to the 3 components of the vector space ($x$, $y$, $z$).  

Substituting then eq.~(\ref{theta-rotated}) into eq.~(\ref{currentECSIM}), we obtain without any approximation or linearization:
\begin{equation}
\overline{\bfJ}_{sg}=\frac{1}{V_g}\sum_p q_p \widehat{\bfv}_p W_{pg} +\frac{\beta_s}{V_g}\sum_p q_p  \widehat{\bfE}_p^{n+\theta} W_{pg}
\label{currentECSIM1}
\end{equation}
where we shortened the notation $W_{pg}=W(\bfx_p^{n+1/2}-\bfx_g)$  and the summation is intended over all particles of species $s$.

We introduce the following hatted current:
\begin{equation}
\widehat{\bJ}_{sg} = \sum_p q_p  \widehat{\bv}_p W_{pg}
\label{hattedmoments}
\end{equation}
with the obvious meaning of current based on the  hatted velocities. 

Using  eq.~(\ref{hatted}) and recalling the definition of eq. (\ref{hattedmoments}), the  expression for the current becomes:
 \begin{equation}
\overline{\bfJ}_{sg}=\widehat{\bJ}_{sg}+\frac{\beta_s}{V_g}\sum_p q_p {\alpha}^{n}_p
\bE_p^{n+\theta}  W_{pg}
\label{currentECSIM2}
\end{equation}
Computing then the electric field on the particles by interpolation form the grid as in eq. (\ref{interpE}), it follows that:
\begin{equation}
\overline{\bfJ}_{sg}=\widehat{\bJ}_{sg}+\frac{\beta_s }{V_g}\sum_p \sum_{g^\prime} q_p {\alpha}^{n}_p
 \bE_{g^\prime}^{n+\theta}  W_{pg^\prime} W_{pg}
\label{currentECSIM3}
\end{equation}
Exchanging the order of summation and introducing the elements of the mass matrices~\cite{burgess1992mass} as:
\begin{equation}
M_{s,gg^\prime}^{ij} = \sum_p q_p {\alpha}^{ij,n}_p W_{pg^\prime} W_{pg}
\label{mass-matrices}
\end{equation}
we obtain, in matrix form:
\begin{equation}
\overline{\bfJ}_{sg}=\widehat{\bJ}_{sg}+\frac{\beta_s }{V_g}\sum_{g^\prime} M_{s,gg^\prime}
 \bE_{g^\prime}^{n+\theta}  
\label{currentECSIMfinal}
\end{equation}
In matrix notation the $3v$ mass matrices are written as $M_{gg^\prime}$, that is without the indices $i,j$ for the vector directions.

Eq. (\ref{mass-matrices}) defines the elements of the mass matrices $M_{s,gg^\prime}$ that are the most peculiar characteristic of the method proposed here. There are  3$v$  such matrices, where $v$ is the dimensionality of the magnetic field and velocity vector, not to be confused with the dimensionality of the geometry used for space $d$. The indices $i$ and $j$ in eq. (\ref{mass-matrices}) vary in the 3$v$-space. For example for full 3-components vectors, $i,j=1,2,3$ and there are 9 mass matrices.  Each matrix is symmetric and very sparse with just $2d$ diagonals.  

Eq. (\ref{currentECSIMfinal}) allows us to retain symmetry between field interpolation to the particles and current interpolation to the grid: with it the computation of the energy exchange between the plasma species and the fields is identical when computed on the particles and when computed on the grid. With this formula we have an explicit linear link between the advanced current at the mid-point of the time step and the electric field at the advanced time. This linear relationship can be substituted into the discretized Maxwell's equations (\ref{maxwell-discrete}) to form a linear set of equations to be solved on the grid:
\begin{equation}
\left\{ \begin{array}{l}
\displaystyle \nabla_g \times \bfE^{n+\theta} + \frac{1}{c} \frac{\bfB^{n+1}-\bfB^n}{\Delta t} =0\\ \\
\displaystyle \nabla_g \times \bfB^{n+\theta} - \frac{1}{c} \frac{\bfE^{n+1}-\bfE^n}{\Delta t} =\frac{4\pi}{c} \left( \widehat{\bJ}_{g}+\sum_{g^\prime} M_{gg^\prime}
 \bE_{g^\prime}^{n+\theta} \right)
\end{array}
\right.
\end{equation}
where we have introduced the total current $\widehat{\bJ}_{g} = \sum_s \widehat{\bJ}_{sg}$ and the species summed mass matrices, that written by elements are:
\begin{equation}
M_{gg^\prime}^{ij} =\sum_s\frac{\beta_s }{V_g}M_{s,gg^\prime}^{ij}
\end{equation}
that can be more usefully held in memory, reducing the memory consumption by a factor equal to the number of species.

The linear relationship in eq. (\ref{currentECSIMfinal})  represents the response of the particles to changes in the fields. Using it in the Maxwell's equations allows one to step over the time scales of wave characteristics if needed, retaining the implicit nature of the scheme. This feature is in stark variance with the fully implicit ECPIC where the full non-linear coupling is kept. The linear response represented by eq. (\ref{currentECSIMfinal}) is sufficient to eliminate the stability requirements. In ECPIC is also possible to use similar linear relationships as preconditions~\cite{chen2015multi}.

This same advantage is also present in all previously known semi-implicit methods. But formula (\ref{currentECSIMfinal}) has one critical new advantage over all other previous semi-implicit methods: it leads to exact energy conservation, just as in the fully implicit ECPIC but without its non-linear complexity. 

Before moving on to the formal proof of the energy conservation, it is important to realize fully the relative complexity of the new ECSIM with respect to the other schemes available.

The particle mover is identical in complexity to the standard Boris scheme of the explicit PIC. In particular it does not need the fixed number of PC iterations used in the IMM  as implemented for example in iPic3D~\cite{ipic3d}. In the typical iPic3D run, 3 steps of PC are used, although other iteration schemes are also possible~\cite{noguchi,peng2015formation}. This new method is then exactly 1/3 in cost in terms of the particle mover step, and by this we include both the motion itself and the field interpolation to the particles. The particle mover is by far the largest 
user of CPU time in the typical cycle. This reduction in cost is  thus very significant.

The field solver is identical to that of other semi-implicit methods. iPic3D for example uses GMRES as solver for the Maxwell's equations. The choice of GMRES is suggested by the non-symmetric nature of the Maxwell matrix and by the diagonal dominance of the matrix which allows a rapid convergence of GMRES. This is in sharp contrast to the solution of elliptic problems, such as the Poisson's equation in electrostatic PIC. In that case, Krylov schemes converge very slowly, requiring the deployment of suitable preconditioner \cite{knoll1999multilevel,kumar2013high}. 
The new ECSIM has one extra cost not present in the standard IMM. The mass matrices need to be evaluated but the extra CPU time and memory is  manageable.

These matrices are computed each time step and used in the solution of the Maxwell equations. The mass matrices are themselves diagonally dominant a feature that retains the diagonal dominance of the overall Maxwell solver leading to a rapid convergence of iterative solvers.

\section{Formal Proof of Energy Conservation}
\label{proof}
Energy conservation means that the amount of energy transferred between the particles and the fields is exactly the same when measured from the particle equations of motions or when measured from the field equations. The result applies only when $\theta=1/2$. Only in that case energy is in fact conserved. A fact we now proceed to prove.

Starting from the particle equations of motions, eq.(\ref{thetaECSIM}), dotting the velocity equation by the average speed, $\overline{\bfv}_p$, and summing over all particles we obtain:
\begin{equation}
\frac{1}{2}\sum_p \left( m_p (\bfv_p^{n+1})^2 - (\bfv_p^{n})^2 \right) =  \Delta t  \sum_p  \left ( q_p  \sum_g \overline{\bfv}_p \cdot\overline{ \bfE}_g W_{pg} \right)
\end{equation}
where the electric field is computed as average consistent with the choice $\theta=1/2$ and the magnetic field drops out as obvious from the properties of the cross product. Exchanging the summation over particles and cells, which is perfectly legitimate for finite sums whitout rising any convergence issue, we obtain:
\begin{equation}
\frac{1}{2}\sum_p \left( m_p (\bfv_p^{n+1})^2 - (\bfv_p^{n})^2 \right) =  \Delta t  \sum_g  \overline{ \bfJ}_g \cdot \overline{ \bfE}_g 
\label{energy-particles}
\end{equation}
where it is recognized that $ \overline{ \bfJ}_g= \sum_p q_p  \overline{\bfv}_p W_{pg} $.

Starting now from the discretized Maxwell equations, eq. (\ref{maxwell-discrete}), and assuming again $\theta=1/2$, we multiply the first equation by $\overline{\bfB}_g$ and the second by $\overline{\bfE}_g$. Summing the two equations it follows that:

\begin{equation}
\begin{array}{l}
\displaystyle\frac{ (\bfB_g^{n+1})^2-(\bfB_g^n)^2 }{2c}+\frac{ (\bfE_g^{n+1})^2-(\bfE_g^n)^2 }{2c} = \\ \\
\displaystyle \Delta t \left( \frac{4\pi}{c} \overline{\bfJ}_g \cdot \overline{\bfE}_g + \bfE_g \cdot \nabla_g \times \bfB - \bfB_g \cdot \nabla_g \times \bfE \right)
\end{array}
\end{equation}

Summing over all grid points and using the properties of the curl operator that are assumed to be replicated by its grid discretization, it follows that:
\begin{equation}
\begin{array}{l}
\displaystyle\sum_g \frac{ (\bfB_g^{n+1})^2-(\bfB_g^n)^2 }{4 \pi}+ \sum_g\frac{ (\bfE_g^{n+1})^2-(\bfE_g^n)^2 }{4\pi} = \\ \\
\displaystyle \Delta t \sum_g\overline{\bfJ}_g \cdot \overline{\bfE}_g +
\frac{c\Delta t}{4 \pi}\sum_g \nabla_g \cdot  (\bfE_g \times \bfB_g )
\end{array}
\label{energy-fields}
\end{equation}

The two terms on the left hand side are recognized as the variation of the magnetic and electric energy as measured on the grid. The first term on the right hand side is the energy exchange term with the particles and the last term expresses the divergence of the Poynting flux.  This is the usual energy equation for electromagnetic energy, discretized on the grid. 

For energy to be conserved in the system, the energy exchange term on the particle equations (eq. (\ref{energy-particles}))  needs to be identical to that on the field equations, (eq. (\ref{energy-fields})). This term is  indeed identically equal to $\sum_g\overline{\bfJ}_g \cdot \overline{\bfE}_g $ in both equations. Energy conservation is enforced exactly, to round off.  

The proof above is based on two key elements. 
First, the current in the field equations, eq. (\ref{energy-particles}), must be computed based on the mean speeds $\overline{\bfv}_p$. This is the guiding principle followed by the fully implicit ECPIC and by the new method presented here. 
Second, the curl operator on the grid must retain the property of the continuum curl operator that $\nabla \cdot (\bfA \times \bfB) = \bfB \cdot \nabla \times \bfA - \bfA \cdot \nabla \times \bfB$. Not all discretizations would have this property but most discretization used in practice have this mimetic property. The discretization used in the example section below does for example. 

This proof just reported is confirmed by the results below that confirm energy is conserved to round off.

\section{Benchmarks}
\label{results}
We have implemented the method described above in two different programs, written in MATLAB. Obviously, production codes based on the new ECSIM formulation will eventually be implemented in more efficient and more easily parallelizable languages. 

Below we report results of test cases from two codes. The first is electrostatic and uses 1 dimension in space, $x$  as well as one dimension in velocity, $v_x$.  The electric field and the current also has one component. In this case, in absence of a magnetic field the mass matrices are all the same and only one mass matrix needs to be computed, instead of 9. This condition is referred to as 1D1V. 

The second program is still one dimensional in space but it has full 3 components of the particle velocity, of the current and of the fields. Obviously, in 1D with the only non-ignorable coordinate being $x$,  the $B_x$ component is constant in time (because the curl operator has no $x$ component) and uniform in space (to satisfy the divergence condition) but not necessarily 0. In this case all 9 mass matrices need to be computed. This second case is referred to as 1D3V. 

In all cases, the ions are a fixed uniform background and only the electrons are followed. All tests are performed on a Apple MacBook Pro with processor Intel Core i7 running at 2.6 GHz and with 16GB of memory.


\subsection{Electrostatic Example: Two-Stream Instability}
A classic benchmark for electrostatic kinetic problems is the two stream instability. The initial condition is formed by a uniform plasma where the particles are distributed in two Maxwellian beams with thermal speed $v_{th}/c=0.01$ and speed $v_0/c=\pm 0.2$ and with equal density $n_0/2$. As mentioned above the system evolution is followed in 1D1V with a system size of $L=2\pi$ and periodic boundary conditions. The analytical solution for the dispersion relation is well known \cite{goldston-rutherford}: longitudinal modes with  $k v_0/\omega_{pe0} < 1$ are unstable and the two beams become destabilized and mix. In the present case, the fastest growing mode has $m=3$ (corresponding to $k=2\pi m/L$) and its growth rate is $\gamma/\omega_{pe}= 0.35$. In the initialization, the particle velocity is displaced by a small ($\Xi=0.1$) amount to favor the onset of the instability: 
$$\delta v_p =  \Xi v_{th}\sin(2\pi m x_p/L)$$

This problem is extremely repeatable and the evolution of the electrostatic energy has been often used in the literature to assess the correctness of a PIC implementation. We compare four runs described below:
\begin{enumerate}
\item The present ECSIM  implemented in a MATLAB code.
\item The fully implicit ECPIC described in \cite{markidis2011energy}, implemented also in MATLAB and thus providing also a direct comparison in terms of language of implementation. The tolerance for the Newton-Krylov iteration is set to $10^{-7}$. 
\item The standard explicit PIC based on the classic textbook  implementation, also implemented in MATLAB for direct performance comparison.
\item The standard electrostatic implementation of the IMM similar to that used in the  implementations of Venus,  Celeste, Parsek2D and iPic3D. Here we use an electrostatic implementation in MATLAB for consistency with the other runs. 
\end{enumerate}

The grid discretization scheme used in all codes above  is summarized in  \ref{spatial-discretization}. All codes use the same time step $\omega_{pe}\Delta t=0.125$ for 1000 cycles and the same number of cells $N=64$ and particles per cell $N_{ppc}=154$.  Such low resolution might give the reader a pause. This is a benchmark and this number of cells has only the purpose of testing the method at low resolution.  We want to assess the ability of energy conserving schemes to work even at low resolution where non-energy conserving codes start to see degraded results. 

\begin{figure}[h]
\begin{center}
\includegraphics[width=\columnwidth]{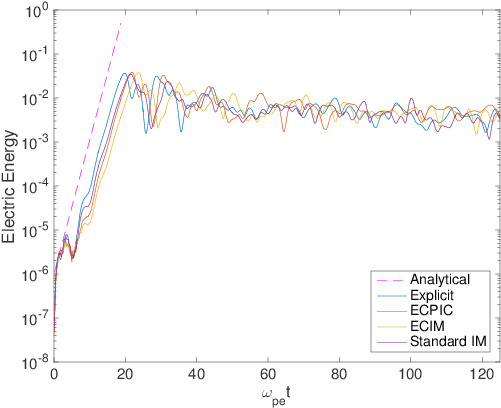}
\caption{Evolution of the integral of the  electric field energy in four runs of the two-stream instability. The linear growth rate predicted by theory is shown by a dashed line. Four runs are compared for the same problem of the two-stream instability: the ECSIM presented here, the standard explicit PIC, the fully implicit ECPIC and the standard IMM. }
\label{2stream-energy}
\end{center}
\end{figure}

\begin{figure}[h]
\begin{center}
\includegraphics[width=\columnwidth]{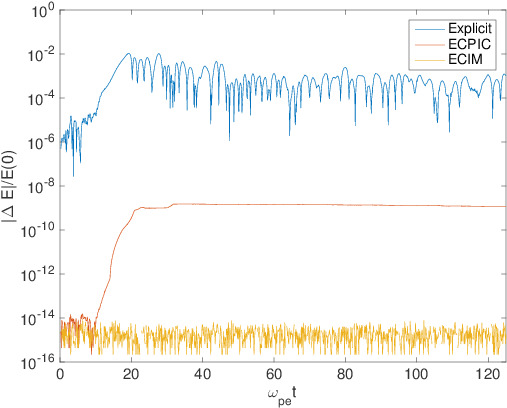}
\caption{Evolution of the error in the  electric field energy in three runs of the two-stream instability: the ECSIM presented here, the standard explicit PIC, the fully implicit ECPIC and the standard IMM. }
\label{2stream-energyerror}
\end{center}
\end{figure}

Figure~\ref{2stream-energy} summarizes all the results from the four different simulations. The results are all similar and all are in good agreement with linear theory.

Figure~\ref{2stream-energyerror} reports the error in the total energy of the same runs. More in detail, Table \ref{2stream-table} reports quantitatively the execution times and energy conservation statistics. 

Energy conservation of the ECPIC and ECSIM is perfect. The fully implicit ECPIC conserves energy to the level of accuracy in the Newton iteration but the ECSIM conserves energy to machine precision. This difference is evident at times $\omega_{pe}t>10$ when the error in the fully energy conserving rises to the level of $10^{-9}$ which corresponds to the allowed tolerance in the Newton iteration.  The difference between the accuracy of the ECPIC and ECSIM is not of any great practical consequence and it is due to the fact that the ECSIM solves the matrix for the Maxwell equation with direct methods that reduce the residual to machine precision. 

The explicit method by comparison has an acceptable energy error of less than 1\%. But 1\% of energy error is still an energy error whose consequences need to be verified in real world simulations to avoid it leading to possible unphysical effects \cite{lapenta2011particle}.

In the present case, the standard implicit moment method fares worse in terms of energy conservation. Using the new ECSIM is especially important in electrostatic cases where the standard IMM  is notoriously challenged.

\begin{table}[h]
\caption{Table of cost effectiveness of Energy conservation in different PIC implementations. Four runs are compared for the same problem of the two-stream instability: the ECSIM presented here, the standard explicit PIC, the fully implicit ECPIC and the standard IMM. }
\begin{center}
\begin{tabular}{|c|c|c|c|}
\hline
Method & CPU time (s) & $\max|\Delta E|/E(0)$ \\ \hline\hline
ECSIM & 3.47 & 8.8057e-15 \\
Explicit PIC & 2.14 & 0.0099\\
EC-PIC & 68.78 & 1.5283e-09\\
Standard IM & 2.43 & 0.0281 \\ 
\hline
\end{tabular}
\end{center}
\label{2stream-table}
\end{table}%

The computation time of the explicit method is smallest. The actual timing observed are just indicative because we made no specific attempt to optimize any of the methods. All methods are implemented in MATLAB using common programming practices appropriate for that language.  Each method can be further optimized and future research can focus on the  best optimization method for the new approach to be compared with the others. 
Especially important is the point that ECPIC can be made much more efficient with preconditioning~\cite{chen2015multi}: the timing comparison provided below is not indicative then of the expected performance of a preconditioned ECPIC.

As noted above the ECSIM is still an explicit method, the additional cost compared with explicit PIC is the computation of two diagonals for the mass matrix. The matrix is symmetric and 3-diagonal so only 2 diagonals need to be computed.  An noted above in absence of magnetic fields there is only one mass matrix rather than 9.
The cost of computing the mass matrix is  proportional to the number of particles and is comparable to that of an interpolation step in the explicit PIC. This cost is not negligible but the tests above show that even with it the computational time does not increase disproportionately.  Future research will need to study the most effective way of coding the mass matrix implementation in modern architectures, using for example vectorization.

\begin{figure}[h]
\begin{center}
\includegraphics[width=\columnwidth]{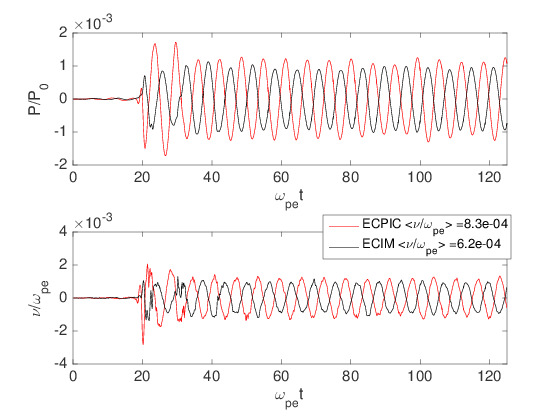}
\caption{Evolution of the fractional change in momentum. The top panel measures the momentum conservation error  as $P/P_0=\sum_p m_p u_p /\sum_p m_p V_0$ where $V_0$ is the initial velocity of the two beams.  The bottom panel measures the error in terms of an effective numerical collision probability defined from the empirical formula $\nu=|dP/dt|/P$. An average collisionality is derived as the standard deviation of the fluctuating value. The fully implicit ECPIC (in back) and the semi implicit ECSIM (in red) are compared.}
\label{2stream-momentum}
\end{center}
\end{figure}

Figure~\ref{2stream-momentum} reports the error in the momentum conservation as a fraction of the initial momentum of each beam. In the ECSIM momentum is not conserved exactly, just as in the full ECPIC and in the IMM. Unfortunately conservation of energy and momentum appear to be mutually exclusive \cite{brackbill2015energy}. In the results reported,  the non conservation of momentum is a tiny fraction of the initial momentum of each beam and it does not accumulate in time. The error is of the order of one part per a thousand for each particle or for the total momentum, since there is no tendency for the error to build up. For the typical applications to space and astrophysical plasmas an error of 1 part per thousand on the momentum is completely negligible. There might be applications where momentum conservation is more important than energy conservation and in that case the explicit momentum conserving method has an advantage.

Another perspective on this is to evaluate the rate of change in the momentum and assess the numerical collisionality that is implicitly leading to that. We can define a numerical collision frequency $\nu$ from:
\begin{equation}
\frac{dP}{dt}=-\nu P
\end{equation}
where $P$ is of the momentum (1D in the present example). This equation needs to be considered with care because the value of $\nu$ is actually oscillating in time and leads to a sort of conservation when  averaged in time, without a diffusive trend leading to an overall decay of momentum. Also in terms of numerical collisionality the effect is of the order of less than one part per thousand without leading to a secular trend. 

For this reason, Figure~\ref{2stream-momentum}  reports the stander deviation of $\nu$ as it oscillates around 0. This quantifies a random walk with no average trend but a standard deviation.

\subsection{Electromagnetic Example: Transverse Beam (Weibel) Instability}

Two counter streaming beams excite also a transverse instability, called Weibel or filamentation, present also in the case of temperature anisotropy~\cite{weibel1959spontaneously, fried}. The instability leads to the excitement of a magnetic field and it is considered one of the fundamental process in Nature (alongside the Bierman battery process) that have allowed the primordial formation of a magnetic field in the Universe. It is a form of dynamo driven by  anisotropic velocity distributions due to temperature anisotropy or beams.

We consider here an initially uniform and unmagnetized plasma in a 1D domain of size $L/d_e=2\pi$ discretized in 64 cells with 154 particles per cell and a time step size of  $\omega_{pe} \Delta t=.125$.  The beams are counterstreaming with speed $v_0/c=0.8$ with a thermal spread of each beam in each direction corresponding to $v_{th0}/c=0.01$. Relativistic effects are moderate at these speeds and we can use the classical methods described above. 

We focus here on a choice of grid and time step where  $c\Delta t>\Delta x$ and exclude  explicit PIC, focusing only on implicit and semi-implicit codes. We use:\begin{enumerate}
\item The present ECSIM method implemented in a MATLAB code for the fully electromagnetic case.
\item The electromagnetic fully implicit EC-PIC described in \cite{markidis2011energy}, implemented also in MATLAB. The tolerance for the Newton-Krylov iteration is set to $10^{-8}$ in the present case. 
\item The standard electromagnetic implementation of the IMM similar to that used in the implementations of Venus,  Celeste, Parsek2D and iPic3D. Here we use a fully electromagnetic implementation in MATLAB for consistency with the other runs.
\end{enumerate}
The codes used here are different from those of the previous electrostatic example, and are fully electromagnetic and 1D3V.  

\begin{figure}[h]
\begin{center}
\includegraphics[width=\columnwidth]{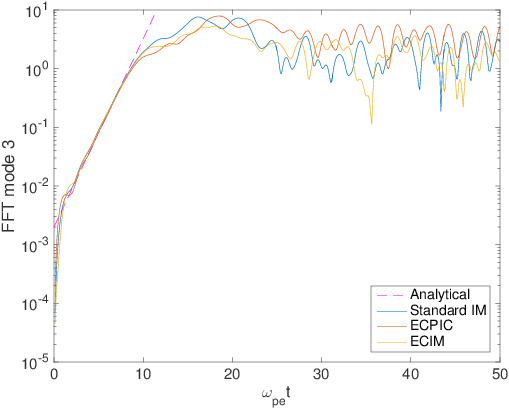}
\caption{Evolution of the amplitude of Fourier mode $m=3$ of the magnetic field component $B_z$  in different runs of the transverse instability. The linear growth rate predicted by theory is shown by a dashed line. Four runs are compared for the same problem of the two-stream instability: the ECSIM presented here, the fully implicit ECPIC and the standard IMM.}
\label{transverse-mode}
\end{center}
\end{figure}

All modes are unstable and we focus especially on the growth of mode $m=3$. Figure \ref{transverse-mode} reports the evolution in time of the mode for different runs. Obviously all runs considered agree with each other and correctly reproduce the growth rate predicted by linear theory.

Again energy conservation is perfect in the ECSIM and is determined by the accuracy of the Newton iteration tolerance in the ECPIC. The standard IMM has a good conservation but not precise. Table \ref{transverse-table} reports the  quantitative information on the energy conservation of the different runs, confirming the qualitative impression of Fig. \ref{transverse-mode}.


 Table \ref{transverse-table} reports also the computational cost of the different runs. Two aspects might appear surprising. First, the fully implicit ECPIC is not much more expensive than the standard IMM. This effect is due to this specific case that does not tax the Newton solver. The standard IMM uses a fixed 3 iteration PC scheme for the particle equations of motion so its cost is not much smaller than that of the Newton iteration in this case. 
 
 The second surprising aspect is that the ECSIM is much less expensive than the standard IMM. The surprise in this is that the ECSIM has the extra cost of computing the 9 mass matrices needed in the scheme. This part costs about 30s of CPU time. But this extra cost not present in the standard IMM is more than compensated for by the fact that the new scheme does not require any PC mover and retains the same numerical cost of a straight through explicit leap-frog scheme. Matrix operations are grid operations and usually in PIC tend to be a minority of the cost, while the lion share is done by the particles.

\begin{figure}[h]
\begin{center}
\includegraphics[width=\columnwidth]{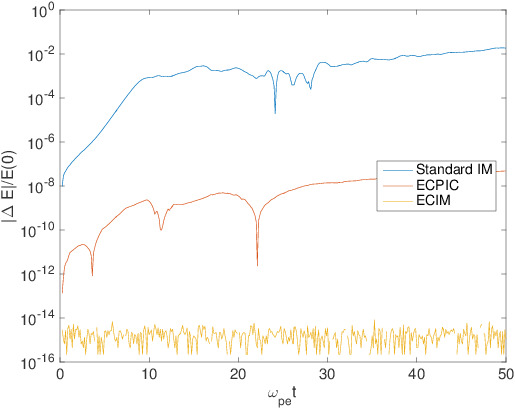}
\caption{Evolution of the error in the  electric field energy in three runs of the transverse instability: the ECSIM presented here, the fully implicit ECPIC and the standard IMM.}
\label{transverse-energy}
\end{center}
\end{figure}

\begin{table}[h]
\caption{Table of cost effectiveness of energy conservation in different PIC implementations. Three runs are compared for the same problem of the transverse instability: the ECSIM presented here, the fully implicit ECPIC~\cite{markidis2011energy} and the standard IMM~\cite{ipic3d}. }
\begin{center}
\begin{tabular}{|c|c|c|c|}
\hline
Method & CPU time (s) & $\max|\Delta E|/E(0)$ \\ \hline\hline
ECSIM & 31.87  & 8.6119e-15\\
EC-PIC & 137.59 & 2.8467e-08\\
Standard IM & 107.28 & 0.0189 \\
\hline
\end{tabular}
\end{center}
\label{transverse-table}
\end{table}%

\section{Elimination of the Finite Grid Instability}
To consider the ability of the ECSIM  to deal with large system sizes, under-resolving the small scales is tested considering a box of size $L=2\pi \Xi$ divided into 64 cells and using 10,000 particles distributed initially uniformly. For small $\Xi$, the grid resolves well the Debye length, $\Delta x/\lambda_{De}=\omega_{pe} L 2 \pi \Xi/V_{th}64$. But as $\Xi$ is increased the grid spacing progressively exceed more and more the stability constraint of explicit PIC: $\Delta x/\lambda_{De}< \xi \pi$. 

\begin{figure}[h]
\begin{center}
\includegraphics[width=\columnwidth]{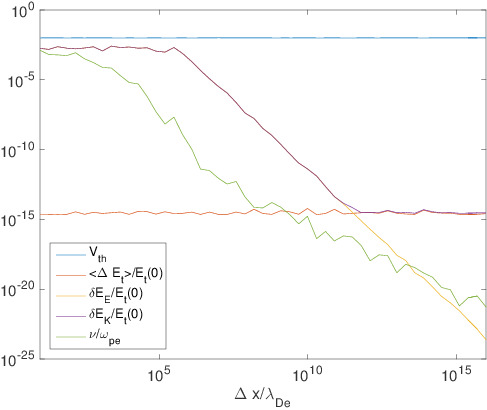}
\caption{Dependence of several run quality factors on the grid spacing normalized to the Debye length of an initially maxwellian and stable plasma. We report 51 different runs with grid spacing ranging over 16 orders of magnitude from $\Delta x/\lambda_{De}=10$ to $\Delta x/\lambda_{De}=10^{16}$. In all runs, $CFL=0.0127$. The parameters reported are: thermal speed at the end of each run,  the average total energy error during each run, normalized to the initial total energy, the standard deviation of the  fluctuation of the electric field, $\delta E_{E}/E_{tot}(0)$ (normalized to the initial total energy), the standard deviation of the fluctuation of the particle kinetic energy, $\delta E_{K}/E_{tot}(0)$ (normalized to the initial total energy) and the effective numerical collision frequency $\nu$ implied by the lack of exact momentum conservation. }
\label{no_finitegrid}
\end{center}
\end{figure}

We consider a uniform plasma with thermal speed $V_{th}=0.01c$ assuming immobile ions. The plasma is in thermal equilibrium and it is stable to all Vlasov-Poisson modes. $\Xi$ is varied by 15 orders of magnitude, from 1 to $10^{15}$, resulting in $\Delta x/\lambda_{De}$  ranging from 10 to $10^{16}$. The time step is selected to $\omega_{pe}\Delta t=.125\Xi$ so that the CFL number is constant in all runs and equal to 0.0127. The results are presented in Fig.~\ref{no_finitegrid}.

\begin{figure}[h]
\begin{center}
\includegraphics[width=\columnwidth]{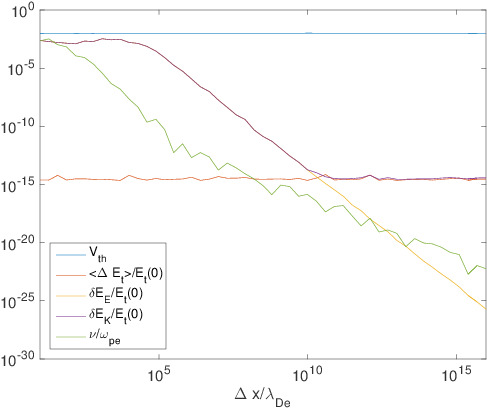}
\caption{Dependence of several run quality factors on the grid spacing normalized to the Debye length of an initially maxwellian and stable plasma. We report 51 different runs with grid spacing ranging over 16 orders of magnitude from $\Delta x/\lambda_{De}=10$ to $\Delta x/\lambda_{De}=10^{16}$. In all runs, $CFL=0.127$. The parameters reported are: thermal speed at the end of each run,  the average total energy error during each run, normalized to the initial total energy, the standard deviation of the  fluctuation of the electric field, $\delta E_{E}/E_{tot}(0)$ (normalized to the initial total energy), the standard deviation of the fluctuation of the particle kinetic energy, $\delta E_{K}/E_{tot}(0)$ (normalized to the initial total energy) and the effective numerical collision frequency $\nu$ implied by the lack of exact momentum conservation. }
\label{no_finitegrid2}
\end{center}
\end{figure}

A second case is selected with $\omega_{pe}\Delta t=1.25\Xi$ with CFL= 0.127. The results of the second ensemble  of runs are presented in Fig.~\ref{no_finitegrid2}. 

In all cases the simulation remains completely stable. The total energy is exactly conserved and the temperature of the electrons remains exactly constant. This result is remarkable for the wide range of scales covered: the method is capable of describing a plasma over 15 orders of magnitude, exceeding the explicit stability limit by 16 orders of magnitude. We stopped at 16, but there is no reason to believe this stable behavior would not continue indefinitely, allowing in principle to model any system size.

The behavior of the fluctuations is significant. As the system size is increased, the thermal fluctuations and the electric field fluctuations are nearly exactly overlapping, as required by the total energy conservation. But at large scales, both fluctuations become progressively smaller. This is the desirable behavior: as the system size becomes larger, the model should transition to a fluid macroscopic behavior where the electric field and thermal energy fluctuations are zero. It happens exactly as one should expect.

The thermal energy has a lower minimum near the machine precision for the linear direct solver, $10^{-15}$ used to solve Maxwell's equaitons. Beyond this level, the thermal energy becomes limited by the total energy error while the electric field fluctuation continues to decrease. The momentum conservation is equally well behaved, with the momentum conservation improving as the system size is increased, another indication of convergence towards a fluid, noiseless, behavior. 

Comparing the two different ensembles of runs with the two different CFL values, the results differ only in detail, presenting the same qualitative behavior. With larger CFL, meaning a larger time step for the same grid spacing, the transition towards reduced fluctuations and convergence towards a macroscopic fluid behavior happens at smaller spatial scales.

\subsection{Artificial Cooling and Artificial Anisotropy}
A peculiar drawback of the lack of energy conservation is the tendency for the particles to heat or cool, anisotropically. 

Explicit codes tend to heat, a tendency that can be especially deleterious providing extra free energy to drive numerical instabilities and to accelerate particles via numerical effects. The terminal stage of this illness being the finite grid instability.

Semi-Implicit codes  tend to cool particles, even though in certain circumstances for large CFL numbers in non-equilibrium situations, numerical heating can be observed as well \cite{directimplicit-optimisation}.

In both cases, the numerical energy error tends to be anisotropic, even though it tends to be gyrotropic. The temperature parallel and perpendicular can change due to numerical effects differently (anisotropy) while the distribution function in the direction perpendicular to the magnetic field remains symmetric (gyrotropic).

The method presented here, as well as the fully implicit methods~\cite{markidis2011energy,chen2011energy} completely eliminate this artificial source of anisotropy. To test this point we consider a condition very typical in the plasma of the Earth space environment (magnetosphere) \cite{lapenta2012particle,lapenta2013space}. The initial electron population is assumed to be uniform and magnetized with $\omega_{ce}=0.005\omega_{pe}$ and $v_{th0}/c=0.01$.  The ions are assumed to be immobile. The magnetic field is directed along $y$  and uniform.

We consider the typical resolution requested in a implicit simulation of the Earth environment: $\Delta x/d_e=0.5$ and $\omega_{ce} \Delta t=0.05$. As a simple test we use a small 1D box with  $L/d_e=32$ and use the 1D3V MATLAB codes described above.

Explicit codes need to resolve the Debye length, but as shown in the previous section, this is not the case of implicit and semi-implicit methods. We assume the electron skin depth needs to be marginally resolved in the case of large magnetospheric simulations where reconnection takes place. The typical scale of the so-called electron diffusion region is of order $d_e$ and needs to be resolved \cite{birn-priest}. An accurate simulation would need to have a smaller grid spacing (see e.g. \cite{lapenta2015secondary}) than the one chosen above, but when one pushes the limit for large scale simulations, the resolution set above would be reasonable \cite{daldorff2014two, ashour2015multiscale, lapenta2015multiscale}.

\begin{figure}[h]
\begin{center}
\includegraphics[width=\columnwidth]{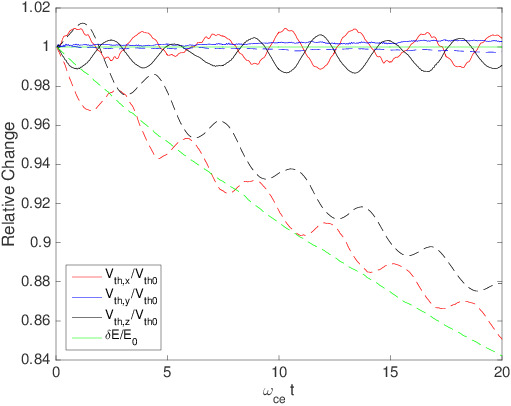}
\caption{Paradigmatic space plasma simulation. Evolution of the total energy and of the electron temperature in each direction for a standard IMM (dashed lines)and for the new ECSIM method (solid lines).  }
\label{cfl}
\end{center}
\end{figure}

Figure~\ref{cfl} reports the comparison of two simulations with the  MATLAB implementations of the standard IMM and of the new ECSIM method. The total energy in the standard IMM method is poorly conserved. In the present case $v_{th0}\Delta t/\Delta x=0.2$ and the IMM is well within its stable range. The explicit method would not work because $\omega_{pe}\Delta t=10$ and $\Delta x/\lambda_{De}=50$, the stability limits of explicit methods widely exceeded. 

During the course of the run, the standard IMM  looses over 10\% of the energy and it  does so all in the perpendicular plane. Eventually the numerical effect would bring the plasma in a regime where instabilities can be driven by the numerical artificial anisotropy. If this were a production run with iPic3D, we would simply have to cut the time step, improving energy conservation. The case selected above is especially pathological and an extreme case of lack of energy conservation not representative of any simulation in the previous literature. We selected it only to make the point clear. The typical work-flow of a iPic3D simulation requires a preliminary study of the optimal $\Delta t$ that allows the longest possible simulation with the largest possible $\Delta t$, while retaining sufficient energy conservation and avoiding artificial anisotropy of the type in Fig.~\ref{cfl}.  This concern is completely eliminated by the new ECSIM making the use of the new ECSIM  simpler. Of course, we have only done a few examples here and only experience will confirm if this initial assessment is correct and it will continue to hold in complex production runs.

\section{Conclusions}
We have introduced a new type of semi-implicit PIC method. The method is  based on a new mover and a new approach to compute the plasma response using the mass matrix instead of the plasma susceptibility  used in previous methods. The complete algorithm derived above is summarized in \ref{summary-algorithm}. 

Let us review what are the novelties and new properties of the new method. 
\begin{enumerate}[I]
\item We use a new mover. The new mover is a hybridization of the  $D_1$ mover of DIM and the $\theta$ mover of IMM. It advances the particle positions just as in the leap-frog mover, stepping the positions over the velocity in a time centered, second order accurate, way. But it advances the velocity using an implicit expression for the electric field, computed as average between start and end of the time step. This feature makes the scheme unconditionally stable and implicit.  

\item We handle the implicitness in the spirit of previous semi-implicit methods but we modify the procedure to linearize the coupling of particles and fields. We do so by introducing the mass matrices that express the particle response to the electric field. The presence of a magnetic field requires to used $3v$ mass matrices, where $v$ is the dimensionality of the fields and velocity vectors. The mass matrices express the plasma response obtained from the particle equations of motion. Once substituted into the Maxwell equations, a complete self consistent linear problem is obtained for the fields. 

\item The new method conserves energy exactly. We provide a direct mathematical proof and we test that in the examples presented energy is indeed conserved to round off. 

\item  The new system is unconditionally stable. The only constraint is that the particles should not move too far in one time step so that the mover remains accurate. The same constraint as in the standard IMM remains valid, $v_{the}\Delta t<\Delta x$. The thermal speed should be changed with the average velocity for a fast moving plasma. This is an accuracy constraint not a stability constraint. With the new approach the current is computed exactly from the mover without the Taylor series expansion typical of the standard IMM. We do not need to worry about the convergence of the truncated series. The condition is purely one of accuracy. In the IMM, instead, violating this condition leads to severe energy error. In the ECSIM, the solution might not be accurate if $v_{the}\Delta t>\Delta x$ but energy would still be exactly conserved.

\item A consequence of the finding above is that the need to resolve the Debye length to avoid the onset of the finite grid instability is completely eliminated. We have shown this point to an exaggerated degree proving stability and accuracy of the solution with cells up to 16 orders of magnitude larger than the Debye length. The ECSIM could model macroscopically large systems with coarse resolution, conserving energy to round off. This results is obtained for a simple maxwellian plasma without drifts: other problems need to be tested to investigate how this property holds up in more complex setups that include drifts. 

\item We have tested the method for accuracy and energy conservation in a few standard benchmark problems, electrostatic and electromagnetic. We reported two here, more have been done but are not reported. 

\item The method presented is exactly energy conserving but it lacks two other conservation properties: momentum and charge. The future users of the method are recommended to monitor if those  quantities are well conserved in realistic applications.

\item The numerical complexity of the ECSIM has been compared with that of the explicit PIC, semi-implicit standard IMM, fully implicit ECPIC. Comparing with explicit PIC, the method has the same complexity in term of particle motion, the most expensive operation for a PIC. But it has the extra cost of the linear solution of the Maxwell equations and the computation of the mass matrices. When compared with standard IMM, the complexity is reduced by a factor of 3 for the particle mover because the ECSIM mover has the same complexity of the leap-frog and does not require the PC iterations of the standard IMM. The Maxwell solver has the same complexity but the ECSIM requires to compute $3v$ extra matrices. Overall we find that the reduced cost of the mover  outweighs the cost of the matrix computation.  To make a full performance comparison with ECPIC, one needs to include the effect of preconditioners~\cite{chen2015multi}. In the present work we have used an unpreconditioned ECPIC but future work will investigate the issue more in depth.

\end{enumerate}

Further research is ongoing to build on the results presented above. First, we are looking at the conservation of other quantities, such as the local divergence of the electric field (Gauss theorem) and that of momentum. Both can be improved if the velocity update is done over substeps \cite{chen2012efficient} rather than in a single jump as in the mover used now (\ref{thetaECSIM}). 

However, the most important future development is implementing the method in a 3D production code. The new method is  sufficiently simple that can be readily implemented in new codes or adapted to existing codes.  Appendix D summarizes all the steps needed for readers interested in coding the new method. The particle mover is very similar to that of explicit PIC and  simpler than in the previous IMM. The field solver can be easily implemented with the now standard GMRES matrix solver.


\section*{Acknowledgments}

The present work is supported by 
the European Commission via the Deep-ER project  (http://www.deep-er.eu), by the 
Onderzoekfonds KU Leuven (Research Fund KU Leuven, GOA scheme and Space Weaves RUN project), by the US Air Force EOARD Project FA2386-14-1-0052 and by the
Interuniversity Attraction Poles Programme of the Belgian Science Policy Office (IAP
P7/08 CHARM). The author is affiliated with Gruppo Nazionale per la Fisica Matematica of the Italian INdAM.

\appendix

\section{Von Neumann Stability Analysis of the New Mover}
\label{stability-mover}

The new  mover of the ECSIM is a modification of the standard $\theta$-mover and it has similar stability properties. We proceed here to show the stability for the paradigmatic case of Langmuir waves used in every textbook analysis of PIC stability~\cite{hockney-eastwood,birdsall-langdon}. 

We proceed using the Von Neumann analysis \cite{isaacson1994analysis}, meaning we assume the temporal evolution to be harmonic so that all quantities depend on time as $e^{i\omega t}$:
\begin{equation}
\begin{array}{c}
\displaystyle \bfx_p^{n+1/2}=\widetilde{\bfx}_p e^{i\omega (n+1/2) \Delta t } \\ \\
\displaystyle \bfv_p^{n+1}=\widetilde{\bfv}_p e^{i\omega  (n+1) \Delta t } \\ \\ 
\displaystyle \bfE^{n+\theta}_p = \widetilde{\bfE}_p \left( (1-\theta) e^{i\omega n \Delta t} +\theta e^{i\omega (n+1) \Delta t} \right)
\end{array}
\end{equation}

We consider  the mover introduced in eq. (\ref{thetaECSIM}) repeated here for the case of a Langmuir wave (i.e. without the magnetic field):
\begin{equation}
\begin{array}{c}
\displaystyle \bfx_p^{n+1/2}=\bfx_p^{n-1/2} + \Delta t \bfv_p^{n}\\ \\
\displaystyle \bfv_p^{n+1}=\bfv_p^{n} + \frac{q_p \Delta t}{m_p}   \bfE_p^{n+\theta} 
\end{array}
\end{equation}
Setting $\theta=1/2$ and substituting in each term the harmonic dependence, we obtain in the Fourier transformed space:
\begin{equation}
\begin{array}{c}
\displaystyle \widetilde{\bfx}_p \left( e^{i\omega \Delta t/2} -e^{-i\omega \Delta t/2}\right) =  \widetilde{\bfv}_p \Delta  t\\ \\
\displaystyle \widetilde{\bfv}_p \left( e^{i\omega \Delta t} -1\right) = \frac{q_s \Delta  t}{m_s} \frac{e^{i\omega \Delta t} +1}{2} \widetilde{\bfE}_p 
\end{array}
\end{equation}

Recalling now the basic property of cold plasma Langmuir waves that relate the variation of the particle displacement with the electric field in Fourier space as \cite{dendy1990plasma}:
 \begin{equation}
\frac{q_s \widetilde{\bfE}}{m_s} = -\omega_{pe}^2 \widetilde{\bfx}
\end{equation}
the final dispersion relation can be computed as the determinant of the system matrix:
\begin{equation}
\begin{array}{c}
\displaystyle \widetilde{\bfx}_p \left( e^{i\omega \Delta t/2} -e^{-i\omega \Delta t/2}\right) =  \widetilde{\bfv}_p \Delta  t\\ \\
\displaystyle \widetilde{\bfv}_p \left( e^{i\omega \Delta t} -1\right) = - \omega_{pe}^2 \Delta t  \frac{e^{i\omega \Delta t} +1}{2} \widetilde{\bfx}_p 
\end{array}
\end{equation}
Using the Euler definition of the trigonometric functions, the determinant is: 
\begin{equation}
\left|
\begin{array}{cc}
\displaystyle 2i \sin \frac{\omega \Delta t} {2}  & \displaystyle -\Delta t \\ \\
\displaystyle \omega_{pe}^2 \cos\frac{\omega \Delta t}{2} &  \displaystyle 2i \sin \frac{\omega \Delta t} {2}
\end{array}
\right| = 0
\end{equation}
Expanding the determinant we obtain the final answer:
\begin{equation}
4 \sin \frac{\omega \Delta t} {2} \tan \frac{\omega \Delta t} {2} = \Delta t^2 \omega_{pe}^2   
\end{equation}
This result differs slightly from that of the $\theta$-scheme that is (also using $\theta=1/2$)~\cite{brackbill-forslund}:
\begin{equation}
4 \tan^2 \frac{\omega \Delta t} {2}  = \Delta t^2 \omega_{pe}^2   
\end{equation}
but remains fundamentally different from the result corresponding to the explicit leapfrog algorithm~\cite{hockney-eastwood}:
\begin{equation}
4 \sin^2 \frac{\omega \Delta t} {2}  = \Delta t^2 \omega_{pe}^2   
\end{equation}

As can be seen, for $\omega_{pe}\Delta t>2$ the leap frog algorithm leads to an impossibility in the real axis, as the $\sin$ function cannot exceed 1 in amplitude. The resulting solutions for $\omega$ are complex conjugate, one resulting in unphysical growth and signaling numerical instability. Both the new scheme and the $\theta$-scheme, instead, have real solutions for $\omega$ for all values of $\Delta t$ and are unconditionally stable.

\begin{figure}
\begin{center}
\includegraphics[width=\columnwidth]{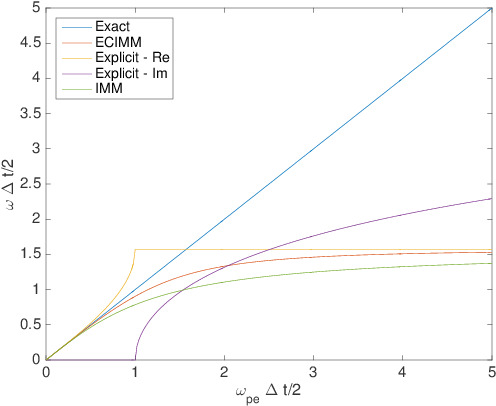}
\caption{Stability analysis for 3 movers: explicit, standard IMM and new ECSIM.  The numerical frequency $\omega$ for Langmuir waves is obtained by the Von Neumann analysis as a function of the time step. }
\label{von-neuman}
\end{center}
\end{figure}

Figure \ref{von-neuman} compares the numerical frequency produced by the three different discretization schemes above. For large $\omega_{pe} \Delta t$, the implicit schemes behave similarly, reaching the asymptotic value of $\omega \Delta t=\pi$, the Nyqvist frequency. The explicit scheme is bound by the constraint  $\omega_{pe} \Delta t<2$ to prevent the numerical frequency from acquiring an imaginary part. 

The two implicit schemes, the new one presented here and the classic $\theta$ scheme differ slightly but the new scheme is more accurate at intermediate $\omega_{pe} \Delta t$, where it remains closer to the physical frequency for the Langmuir waves. The slight improvement is unlikely to have great practical consequences. Nevertheless, it is better to be better, even if slightly.

\section{Spatial Discretization and its Properties}
\label{spatial-discretization}

The main paper uses a shorthand notation for the grid discretization: $\nabla_g \times \bfE$ for example. The results reported in the present paper do not depend critically on the spatial discretization, only the temporal discretization counts. Nevertheless, for the final implementation in the result section we need to use a specific discretization. We use the same type of discretization used in the family of IMM codes: Venus, Celeste, Parsek2D and iPic3D. The grid is formed by cells, the centers of the cells host the magnetic field and the vertices of the cells host the electric field. 

In the example section we apply this in 1D. The application is straightforward elementary numerical analysis. Here we very briefly give the details. An interval from $x \in [0,L]$ with periodic conditions is discretized in $N$ cells. The generic cell $i$ is an interval $[x_{i},x_{i+1}]$ with center $x_{i+1/2}$. We use a uniform gird, so $x_i = i \Delta x$ with the integer $i \in [0,N]$. 

The electric field is then located at $\bfE(x_i)=\bfE_i$ and the magnetic field at $\bfB(x_{i+1/2})=\bfB_{i+1/2}$. The shorthand notation  $\nabla_g \times \bfE$ can then be given a concrete 
meaning:
\begin{equation}
(\nabla \times \bfE)_{i+1/2}= \frac{1}{\Delta x}
\left(\begin{array}{c}
0\\ -(E_{z,i+1}-E_{z,i})\\(E_{y,i+1}-E_{y,i})
\end{array}
\right)
\end{equation}
Analogously, for the curl of the magnetic field, $\nabla_g \times \bfB$,  we have:
\begin{equation}
(\nabla \times \bfB)_{i}= \frac{1}{\Delta x}
\left(\begin{array}{c}
0\\ -(B_{z,i+1/2}-B_{z,i-1/2})\\(B_{y,i+1/2}-B_{y,i-1/2})
\end{array}
\right)
\end{equation}

These expression can be used to explicitly write out the Maxwell's equations. It is interesting to express the Poynting flux term in the energy equation:
\begin{equation}
\begin{array}{c}
\displaystyle \sum_g V_g (\bfE_g \cdot \nabla_g \times \bfB - \bfB_g \cdot \nabla_g \times \bfE) \rightarrow \\ \\ 
\displaystyle \sum_i \left( 
B_{y,i+1/2}(E_{z,i+1}-E_{z,i}) - B_{z,i+1/2} (E_{y,i+1}-E_{y,i}) \right. \\
\displaystyle \left. -E_{y,i}(B_{z,i+1/2}-B_{z,i-1/2}) +E_{z,i}(B_{y,i+1/2}-B_{y,i-1/2}) \right) = 0
\end{array}
\end{equation}
All  terms $i \in [2,N-1]$  cancel out, leaving only the boundary values corresponding to the end values $i=0$ and  $i=N$. Via the periodic boundary conditions these terms are the same and cancel out as well. In presence of periodic boundary conditions the Poynting flux terms gives an exactly 0 balance as physically meaningful.

\section{Rate of Convergence in Space and Time of the ECSIM}
A powerful tool to detect coding errors \cite{roache2002code} is to test the actual observed rate of convergence and compare it with the order of accuracy theoretically predicted by the modified equation approach~\cite{warming1974modified}. We apply the approach to each module of the code separately:  the field equations and the mover. We do not make a full coupling analysis because it depends also on the number of particles and their shape. We analyse the Maxwell solver with given sources and the mover with a given electric fields (that of linear Langmuir waves), in absence of magnetic fields. Note that the coupling with the magnetic field is instead first order in time.

The spatial discretization used for the Maxwell equations is centered and produces a second order accurate convergence. The temporal discretization of the Maxwell equations and of the equations of motion is also centered (in the case of $\theta=1/2$) and leads also to a second order accurate scheme. 

Except for coding error, the rate of convergence obtained solving the method described above with progressively smaller $\Delta t$ and $\Delta x$ should therefore also be second order.

\begin{figure}
\begin{center}
\begin{tabular}{cc}
\includegraphics[width=0.45\columnwidth]{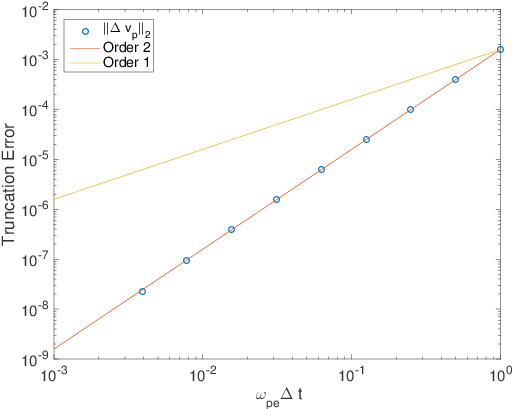}&
\includegraphics[width=0.45\columnwidth]{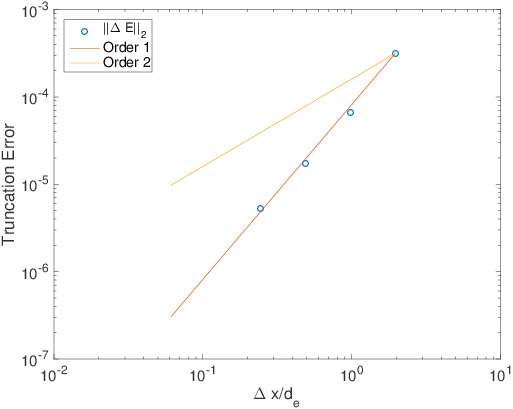}
\end{tabular}
\caption{Stability analysis for 3 movers: explicit, standard IMM and new ECSIM.  The numerical frequency $\omega$ for Langmuir waves is obtained by the Von Neumann analysis as a function of the time step. }
\label{convergence}
\end{center}
\end{figure}

Figure \ref{convergence} confirms that the discretization error  in the mover decreases with the second power of both $\Delta t$ and $\Delta x$. The results are obtained decreasing $\Delta t$ at fixed  $\Delta x$ (left panel) and viceversa (right panel). The error is measured in $L_2$ norm with respect to a converged solution.

The combined confirmation of exact energy conservation and second order accuracy in space and time can be used as powerful debugging tool. We have experienced that small errors even just in the boundary conditions break energy conservation. 

\section{Summary of the ECSIM Algorithm}
\label{summary-algorithm}
We summarize below the steps needed in the construction of the ECSIM algorithm
\begin{enumerate}
\item Advance the particle position based on the know velocity, as in the leap-frog scheme of explicit PIC:
\begin{equation}
\bfx_p^{n+1/2}=\bfx_p^{n-1/2} + \Delta t \bfv_p^{n}
\end{equation}
\item Interpolate the known magnetic field to the particles:
\begin{equation}
\bfB^n_p(\bfx_p)=\sum_g \bfB^n_g W_{pg}
\end{equation}
where $W_{pg}=W(\bfx_p^{n+1/2}-\bfx_g)$.
\item Compute the rotation operators. This can be an explicit matrix in the code or it can be a method (in the C++ sense, a routine in Fortran) for matrix-free formulations:
\begin{equation}
{\alpha}_p^n =  \frac{1}{1+(\beta_s B_p^{n})^2}
\left(\mathbb{I}-\beta_s \mathbb{I} \times \bB_p^n +\beta_s^2
\bB_p^n \bB_p^n \right)
\end{equation}
where $\beta_s=q_p \Delta t/2m_p$.
\item Compute the hatted current based on the known velocities rotated by the known magnetic field:
\begin{equation}
\widehat{\bJ}_{sg} = \sum_p q_p  {\alpha}^n_p  \bv^n_p W_{pg}
\end{equation}
with the total current $\widehat{\bJ}_g =\sum_s \widehat{\bJ}_s $.
\item Construct the mass matrices, $3v$ in number, where $v$ is the dimensionality of the vectors (not to be confused with the dimensionality of space $d$). Again this can be a matrix stored in memory or a method for matrix free formulations. The elements of the matrices are:
\begin{equation}
M_{gg^\prime}^{ij} = \sum_s \left( \frac{\beta_s }{V_g} \sum_p q_p {\alpha}^{ij,n}_p W_{pg^\prime} W_{pg}\right)
\end{equation}
where $V_g$ is the volume  of cell $g$. In matrix notation the $3v$ mass matrices are writen as $M_{gg^\prime}$, that is without the indeces $i,j$ for the vector directions.
\item Solve the linear set of Maxwell equations where the plasma response is provided by the mass matrices:
\begin{equation}
\left\{ \begin{array}{l}
\displaystyle \nabla_g \times \bfE^{n+\theta} + \frac{1}{c} \frac{\bfB^{n+1}-\bfB^n}{\Delta t} =0\\ \\
\displaystyle \nabla_g \times \bfB^{n+\theta} - \frac{1}{c} \frac{\bfE^{n+1}-\bfE^n}{\Delta t} =\frac{4\pi}{c} \left( \widehat{\bJ}_{g}+\sum_{g^\prime} M_{s,gg^\prime}
 \bE_{g^\prime}^{n+\theta} \right)
\end{array}
\right.
\end{equation}
\item Interpolate the new field from the grid to the particles:
\begin{equation}
\bfE_p^{n+\theta}(\bfx_p)=\sum_g \bfE_g^{n+\theta} W_{pg}
\end{equation}
\item Advance the particle velocity using the new electric field:
\begin{equation}
\overline{\bfv}_p={\alpha}^n_p  \left( \bv^n_p+
\beta_s {\bE}_p^{n+\theta}(\bx_p^{n+1/2})  \right)
\end{equation}
The new velocity is then $\bfv_p^{n+1} =2 \overline{\bfv}_p -\bfv_p^{n}$.
\end{enumerate}

As can be observed the computational cycle retains the same nature as that of explicit PIC or of other semi-implicit PIC. There is no iteration between particles and field.



\bibliographystyle{elsarticle-num}
\bibliography{bibliografie}







\end{document}